\RequirePackage{lineno}
\documentclass[aps,prd,preprintnumbers,amsmath,amssymb,twocolumn,nofootinbib]{revtex4-1}
\usepackage{graphicx}
\usepackage{xspace}
\usepackage{siunitx}
\usepackage[utf8]{inputenc}
\usepackage{dcolumn}

\newcommand{\rmu}{R_\mu}
\newcommand{\mrmu}{\langle \rmu \rangle}
\newcommand{\lnrmu}{\ln\!\rmu}
\newcommand{\mlnrmu}{\langle \lnrmu \rangle}
\newcommand{\lne}{\ln\!E}
\newcommand{\lna}{\ln\!A}
\newcommand{\nmu}{N_\mu}

\newcommand{\lnnmu}{\ln\!\nmu}
\newcommand{\mlnnmu}{\langle \ln\!\nmu \rangle}
\newcommand{\mlna}{\langle \lna \rangle}
\newcommand{\xmax}{X_\text{max}}
\newcommand{\mxmax}{\langle \xmax \rangle}
\newcommand{\rmumc}{R_\mu^\text{MC}}
\newcommand{\nnt}{N_{19}}
\newcommand{\fg}[1]{Fig.\,\ref{fig:#1}\xspace}
\newcommand{\eq}[1]{Eq.\,(\ref{eq:#1})\xspace}
\newcommand{\corsika}{\textsc{Corsika}}
\newcommand{\aires}{\textsc{Aires}}
\newcommand{\qgsjet}{\textsc{QGSJet}}
\newcommand{\epos}{\textsc{Epos}}
\newcommand{\geant}{\textsc{Geant}}

\newcommand{\de}{\text{d}}
\newcommand{\loggain}{\de \mlnrmu / \de \lne}
\newcommand{\sys}{\text{(sys.)}}
\newcommand{\ts}{,} 

\begin{document}

\preprint{Published in Physical Review D as doi:10.1103/PhysRevD.91.032003}


\title{Muons in air showers at the Pierre Auger Observatory: Mean number in highly inclined events}

\author{
\par\noindent
{\bf The Pierre Auger Collaboration} \\
A.~Aab$^{42}$, 
P.~Abreu$^{64}$, 
M.~Aglietta$^{53}$, 
E.J.~Ahn$^{81}$, 
I.~Al Samarai$^{29}$, 
I.F.M.~Albuquerque$^{17}$, 
I.~Allekotte$^{1}$, 
J.~Allen$^{85}$, 
P.~Allison$^{87}$, 
A.~Almela$^{11,\: 8}$
% ...

\par\noindent
$^{1}$ Centro At\'{o}mico Bariloche and Instituto Balseiro (CNEA-UNCuyo-CONICET), San 
Carlos de Bariloche, 
Argentina \\
% ...
$^{8}$ Instituto de Tecnolog\'{\i}as en Detecci\'{o}n y Astropart\'{\i}culas (CNEA, CONICET, UNSAM), 
Buenos Aires, 
Argentina \\
% ...
$^{11}$ Universidad Tecnol\'{o}gica Nacional - Facultad Regional Buenos Aires, Buenos Aires,
Argentina \\
% ...
$^{17}$ Universidade de S\~{a}o Paulo, Instituto de F\'{\i}sica, S\~{a}o Paulo, SP, 
Brazil \\
% ...
$^{29}$ Institut de Physique Nucl\'{e}aire d'Orsay (IPNO), Universit\'{e} Paris 11, CNRS-IN2P3, 
Orsay, 
France \\
% ...
$^{42}$ Universit\"{a}t Siegen, Siegen, 
Germany \\
% ...
$^{53}$ Osservatorio Astrofisico di Torino  (INAF), Universit\`{a} di Torino and Sezione INFN, 
Torino, 
Italy \\
% ...
$^{64}$ Laborat\'{o}rio de Instrumenta\c{c}\~{a}o e F\'{\i}sica Experimental de Part\'{\i}culas - LIP and  
Instituto Superior T\'{e}cnico - IST, Universidade de Lisboa - UL, 
Portugal \\
% ...
$^{81}$ Fermilab, Batavia,  IL 60510-0500, 
USA \\
% ...
$^{85}$ New York University, New York, NY 10003, 
USA \\
% ...
$^{87}$ Ohio State University, Columbus, OH 43210-1061, 
USA \\
% ...
\par\noindent
(\ddag) Deceased \\
(a) Now at Konan University \\
(b) Also at the Universidad Autonoma de Chiapas on leave of absence from Cinvestav \\
(c) Now at NYU Abu Dhabi \\
}

\begin{abstract}

We present the first hybrid measurement of the average muon number in air showers at ultra-high energies, initiated by cosmic rays with zenith angles between $62^\circ$ and $80^\circ$. The measurement is based on 174 hybrid events recorded simultaneously with the Surface Detector array and the Fluorescence Detector of the Pierre Auger Observatory. The muon number for each shower is derived by scaling a simulated reference profile of the lateral muon density distribution at the ground until it fits the data. A $10^{19}$~eV shower with a zenith angle of $67^\circ$, which arrives at the Surface Detector array at an altitude of \SI{1450}{m} above sea level, contains on average $(2.68 \pm 0.04 \pm 0.48\,\sys) \times 10^{7}$ muons with energies larger than \SI{0.3}{GeV}. The logarithmic gain $\text{d}\ln{N_\mu} / \text{d}\ln{E}$ of muons with increasing energy between \SI{4e18}{eV} and \SI{5e19}{eV} is measured to be $(1.029\, \pm\, 0.024\, \pm 0.030\,\sys)$.

\end{abstract}

\maketitle


\section{Introduction}
Understanding the mass composition of ultra-high energy cosmic rays at Earth is fundamental to unveil their production and propagation mechanisms. The interpretation of observed anisotropies~\cite{Abraham2007, Abreu2010} and of features in the flux relies on it, such as the break in the power law spectrum around \SI{4e18}{eV}, and the flux suppression above \SI{4e19}{eV}~\cite{Abraham2010a}.

Ultra-high energy cosmic rays can only be observed indirectly through air showers. The mass composition of cosmic rays can be derived from certain air shower observables, but the inference is limited by our theoretical understanding of the air shower development~\cite{Abreu2013}. Air shower simulations require knowledge of hadronic interaction properties at very high energies and in phase space regions that are not well covered by accelerator experiments. The systematic uncertainty of the inferred mass composition can be reduced by studying different observables (see, e.g., \cite{Kampert2012}). The slant depth $\xmax$ of the shower maximum is a prominent mass-sensitive tracer, since it can be measured directly with fluorescence telescopes.

The number of muons in an air shower is another powerful tracer of the mass. Simulations show that the produced number of muons, $\nmu$, rises almost linearly with the cosmic-ray energy $E$, and increases with a small power of the cosmic-ray mass $A$. This behavior can be understood in terms of the generalized Heitler model of hadronic air showers~\cite{Matthews2005}, which predicts
\begin{linenomath}
\begin{equation}\label{eq:nmu_heitler}
\nmu = A \left(\frac{E/A}{\xi_{\mathrm c}}\right)^\beta,
\end{equation}
\end{linenomath}
where $\xi_{\mathrm c}$ is the critical energy at which charged pions decay into muons and $\beta \approx 0.9$. Detailed simulations show further dependencies on hadronic-interaction properties, like the multiplicity, the charge ratio and the baryon anti-baryon pair production~\cite{Pierog2008, Ulrich2011a}.

To use the muon number $\nmu$ as a tracer for the mass $A$, the cosmic-ray energy $E$ has to be independently measured event-by-event with a small systematic uncertainty. By taking the logarithm of \eq{nmu_heitler} and computing the derivative, we obtain the logarithmic gain of muons with increasing energy
\begin{linenomath}
\begin{equation}\label{eq:slope_heitler}
  \frac{\de \lnnmu}{\de \lne} = \beta + (1-\beta) \frac{\de \lna}{\de \lne},
\end{equation}
\end{linenomath}
which carries additional information on the changes in the mass composition and is invariant to systematic offsets in the energy scale. The dependency of the muon number $\nmu$ on the mass of cosmic rays is complementary to other mass-sensitive observables such as the depth of the shower maximum, $\xmax$. If both observables are combined, the internal consistency of hadronic interaction models can be tested.

We present the average number of muons in inclined showers above \SI{4e18}{eV} measured with the Pierre Auger Observatory~\cite{Abraham2004}, which is located in Mendoza province, Argentina. The Pierre Auger Observatory was completed in 2008 and covers an area of 3000\,\si{km^2}. It is a hybrid instrument to detect cosmic-ray induced air showers, which combines a Surface Detector array (SD) of 1660 water-Cherenkov stations~\cite{Abraham2010b} placed on a triangular grid with \SI{1.5}{km} spacing, with a Fluorescence Detector (FD)~\cite{Abraham2010c}. Due to their cylindrical volume, the Surface Detectors are sensitive to inclined and even horizontal particles~\cite{Abraham2007b,Abraham2009b}. On dark nights, which corresponds to a duty cycle of about 13\,\%, the longitudinal shower development and the calorimetric energy of the shower are measured by the FD. It consist of 27 telescopes with UV-filters located at four sites around the SD array, each monitoring a $30^{\circ}\times 28^{\circ}$ patch of the sky.

Extensive air showers with zenith angles exceeding $62^\circ$ are characterized at the ground by the dominance of secondary energetic muons, since the electromagnetic component has been largely absorbed in the large atmospheric depth crossed by the shower. Such \emph{inclined showers} provide a direct measurement of the muon number at the ground~\cite{Ave2003a}. The muon number in less inclined air showers has also been explored~\cite{Kegl2013,Farrar2013}, but the measurement is in this case complicated by the need to separate the electromagnetic and the muonic signals in surface detectors. The unique features of showers around $60^\circ$ zenith angle further led to the derivation of the muon production depth (MPD) from the arrival times of signals in the SD~\cite{Aab2014}, which is another powerful observable to study the mass composition and hadronic interaction models.

We measure the muon number in inclined air showers using the relative scale factor $\nnt$ which relates the observed muon densities at the ground to the average muon density profile of simulated proton-induced air showers of fixed energy $10^{19}\,\si{eV}$. This approach follows from developments that have been introduced to reconstruct inclined showers, taking into account the rich spatial structure of the muon distributions at the ground. The scale factor $\nnt$ is independent of the zenith angle and details of the location of the observatory~\cite{Ave2000b,Ines2013} and can be also used as an estimator of the muon number. These developments led to the first limit on the fraction of cosmic photons in the EeV energy range~\cite{Ave2000} and to an independent measurement of the energy spectrum of cosmic rays~\cite{Dembinski2011a}.

\section{Reconstruction of the muon number}
Inclined showers generate asymmetric and elongated signal patterns in the SD array with narrow pulses in time, typical for a muonic shower front. Events are selected by demanding space-time coincidences of the signals of triggered surface detectors which must be consistent with the arrival of a shower front~\cite{Abraham2010b,Vazquez2009}. After event selection, the arrival direction $(\theta, \phi)$ of the cosmic ray is determined from the arrival times of this front at the triggered stations by fitting a model of the shower front propagation. The achieved angular resolution is better than $0.6^\circ$ above \SI{4e18}{eV}~\cite{Rodriguez2011}.

%
%
\begin{figure}[t]
\centering
\includegraphics[width=0.47\textwidth]{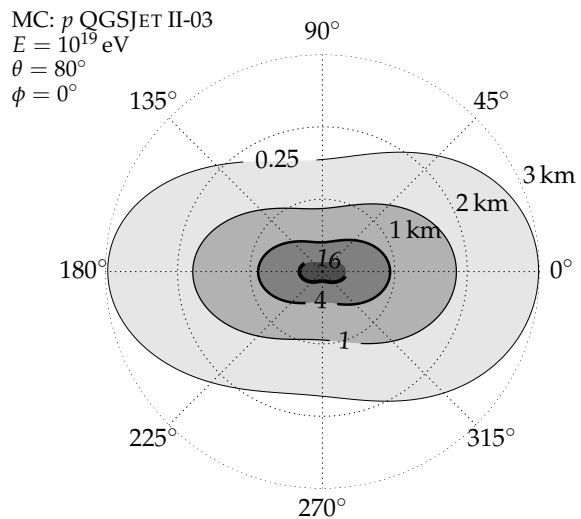}
\caption{Expected number of muon hits per SD station as predicted by the reference profile $\rho_{\mu,19}$, for $\theta=80^\circ$ and $\phi=0^\circ$, in cylindrical coordinates around the shower axis. The radial density roughly follows a power law in any given direction. The quadrupole structure is generated by charge separation in Earth's magnetic field. The weaker dipole structure is caused by projection effects and muon attenuation. Early (late) arriving particles are on the right (left) side in this projection.}
\label{fig:fig01}
\end{figure}
%
%

Once the shower direction is established, we model the muon density $\rho_\mu$ at the ground point $\vec r$ as
\begin{linenomath}
\begin{equation}\label{eq:nmudef}
\rho_\mu(\vec r) = \nnt \; \rho_{\mu,19}(\vec r; \theta, \phi),
\end{equation}
\end{linenomath}
where $\rho_{\mu,19}$ is the parametrized ground density for a proton shower simulated at $10^{19}$\,\si{eV} with the hadronic interaction model \qgsjet II-03~\cite{Ostapchenko2006}. An example is given in \fg{fig01}. It was shown in detailed studies~\cite{Ave2000a,Dembinski2010} that the attenuation and shape of $\rho_{\mu,19}$ depend very weakly on the cosmic-ray energy $E$ and mass $A$ for showers with $\theta > 60^\circ$, so the factorization in \eq{nmudef} is a good approximation for showers above $10^{18}$\,\si{eV}. It was also shown that the lateral shape of $\rho_{\mu,19}$ is consistently reproduced by different hadronic interaction models and air-shower simulation codes. The lateral shape at the ground is mainly determined by hadronic interactions at beam energies of up to a few hundred GeV, in which models are constrained by data from fixed target experiments. The strong zenith angle dependence is factorized out into $\rho_{\mu,19}$ in \eq{nmudef}, so that the scale factor $\nnt$ at a given zenith angle is a relative measure of the produced number of muons $N_\mu$, addressed in \eq{nmu_heitler}.

The scale factor $\nnt$ is inferred from measured signals with a maximum-likelihood method based on a probabilistic model of the detector response to muon hits obtained from \geant4~\cite{Allison2006} simulations with the Auger Offline software framework~\cite{Argiro2007}. A residual electromagnetic signal component is taken into account based on model predictions (typically amounting to $20\,\%$ of the muon signal)~\cite{Valino2010}. The procedure is described in full detail in Ref.~\cite{HasReconstructionJCAP}.

The reconstruction approach was validated in an end-to-end test with three sets of simulated events. The first set consists of 100\ts000 proton and 100\ts000 iron showers generated with \aires~\cite{Sciutto2001}, using \qgsjet01~\cite{Kalmykov1993}. Showers following an $E^{-2.6}$ energy spectrum and an isotropic angular distribution were simulated at a relative thinning of $10^{-6}$. The second (third) set consists of 12\ts000 proton and 12\ts000 iron showers generated using \corsika~\cite{Heck1998}, with \qgsjet II-04~\cite{Ostapchenko2011} (\epos\ LHC~\cite{Pierog2013}), with the same thinning and angular distribution and an $E^{-1}$ energy spectrum. Showers have subsequently undergone a full simulation of the detector, with random placement of impact points in the SD array. Simulated and real events were reconstructed with the same procedure.

%
%
\begin{figure}[t]
\centering
\includegraphics[width=0.47\textwidth,keepaspectratio,clip]{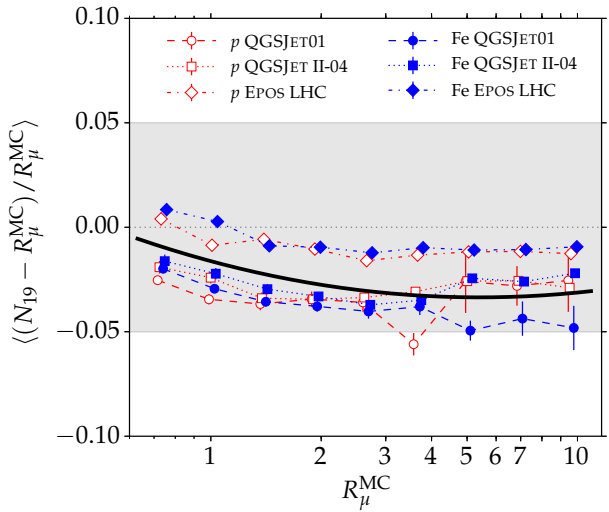}
\caption{Average relative deviation of reconstructed muon content $\nnt$ from the true muon content $\rmumc$ (as defined in the text) for proton and iron showers. The shaded area indicates the systematic uncertainty of $\nnt$. The solid black line represents a second order polynomial adjusted to describe the mean bias.}
\label{fig:fig02}
\end{figure}
%
%

For each MC event we compute the ratio $\rmumc = \nmu/N_{\mu,19}$ by counting the total number of muons $\nmu$ at the ground in the simulation and dividing by the total number of muons $N_{\mu,19} = \int \text{d}y \int\rho_{\mu,19}\,\text{d}x$ obtained by integrating the reference model. We compare this ratio with the value of $\nnt$ obtained from the fit of \eq{nmudef}.

The relative deviation of $\nnt$ from $\rmumc$ is shown in \fg{fig02} to be within $5\,\%$ for events with $\rmumc > 0.6$. This confirms the factorization hypothesis of \eq{nmudef}, the approximate universality of the chosen reference profile, and validates the reconstruction method. The largest source of systematic bias is the remaining model dependence of the reference profile $\rho_{\mu,19}(\vec r)$. To get an unbiased estimator, we correct the measured value $\nnt$ for the average bias. We use a second order polynomial as indicated in \fg{fig02} to reproduce $\rmumc$ to within 3\,\% for the latest models. We consequently call the corrected estimator $\rmu$ in the following.

We constructed in this way an unbiased estimator of the total number of muons at the ground that is nearly independent of model assumptions and the zenith angle of the shower. The value $\rmu = 1$ corresponds to \num{2.148e7} (\num{1.202e+07}, \num{5.223e+06}) muons with energies above \SI{0.3}{GeV} (Cherenkov threshold for muons in water) that reach the Auger site at an altitude\footnote{Altitudes are given with respect to the WGS~84 reference ellipsoid~\cite{WGS84}.} of \SI{1425}{m} at a shower inclination of $60^\circ$ ($70^\circ$, $80^\circ$). By combining the model uncertainty with that of the simulated muon response of the detectors at $\theta > 60^\circ$~\cite{Ghia2007}, we conservatively estimate the systematic uncertainty of $\rmu$ to be 11\,\%.

\section{Data set and event selection}
We proceed to study the muon content $\rmu$ of inclined showers as a function of the cosmic-ray energy $E$. The calorimetic energy $E_\text{cal}$ of the shower is measured independently by the FD in a subset of hybrid events recorded simultaneously in FD and SD. The total energy $E$ is computed by adding the average invisible energy $\langle E_\text{inv} \rangle$, which has been re-evaluated recently based on data~\cite{Verzi2013}. Since $\rmu$ is sensitive to the cosmic-ray mass $A$, we make sure to not to bias the selected sample towards certain masses by a careful selection of the accepted shower geometries.

The data set consists of hybrid events with zenith angles $62^\circ < \theta < 80^\circ$ and at least four triggered stations. Only events well-contained in the SD array are considered; the station closest to the fitted core and its six adjacent stations need all to be active. The FD measurements have to pass quality cuts designed to ensure an accurate reconstruction of arrival direction and longitudinal profile. The cuts are adapted versions of those used in calibration of events with $\theta < 60^\circ$~\cite{Pesce2011}. The SD station used in the FD geometrical reconstruction must be closer to the core than \SI{750}{m}. Only events with good atmospheric conditions are considered: the vertical aerosol optical depth needs to be measured and has to be less than 0.1; if cloud information is available we require a cloud coverage below 25\,\% in the field of view, a distance from the cloud layer to the measured profile larger than \SI{50}{g cm^{-2}}, and a thickness of the cloud layer less than \SI{100}{g cm^{-2}}. The few remaining longitudinal profiles affected by clouds are rejected by requiring a small $\chi^2$-residual in the Gaisser-Hillas fit, $(\chi^2 - n_\text{dof})/\sqrt{2n_\text{dof}} < 3$, and the parameter $X_0$ of the fitted Gaisser-Hillas profile must be negative.

In addition to the quality selection criteria, a fiducial cut on the FD field of view is applied to ensure that it is large enough to observe the depth of shower maximum with equal probability within the range of plausible values. This cut also ensures a maximum accepted uncertainty of the depth of the shower maximum of \SI{150}{g cm^{-2}}, and a minimum viewing angle of light in the FD telescope of $25^\circ$. Finally, we accept only energies above \SI{4e18}{eV} to ensure a trigger probability of 100\,\% for FD and SD.

The selection is applied to inclined events recorded from 1 January 2004 to 1 January 2013. Out of 29722 hybrid events, 174~events are accepted. Due to the geometrical acceptance of the SD and the fiducial cut on the FD field of view, the zenith angle distribution peaks near $62^\circ$. The average zenith angle is $(66.9 \pm 0.3)^\circ$ and the highest energy in the sample is $(48.7 \pm 2.9) \times 10^{18}\,\si{eV}$.

\section{Data analysis}
The muon content $\rmu$ of individual showers with the same energy $E$ and arrival direction varies. This is caused by statistical fluctuations in the development of the hadronic cascade, and, in addition, by random sampling from a possibly mixed mass composition. We will refer to these fluctuations combined as intrinsic fluctuations. In the following, we will make statements about the average shower, meaning that the average is taken over these intrinsic fluctuations. Detector sampling adds Gaussian fluctuations to the observed value of $\rmu$ on top of that. The statistical uncertainties of $\rmu$ and $E$ caused by the sampling are estimated by the reconstruction algorithms event-by-event. We will refer to them as detection uncertainties.

From \eq{nmu_heitler} we expect that the average number of produced muons, which is proportional to $\mrmu$, and the cosmic-ray energy $E$ have a relationship that is not far from a power law. Therefore we fit the parametrization
\begin{linenomath}
\begin{equation}\label{eq:fit}
\mrmu = a\, (E/10^{19}\,\text{eV})^b
\end{equation}
\end{linenomath}
to the selected data set, using a detailed maximum-likelihood method that takes the mentioned fluctuations into account. Intrinsic fluctuations of $\rmu$ are modeled with a normal distribution that has a constant relative standard deviation $\sigma[\rmu]/\rmu$. This model is found to be in good agreement with shower simulations. The $a$ parameter of the fitted curve represents the average muon content $\mrmu (10^{19}\,\text{eV})$ at $10^{19}$\,eV, and the $b$ parameter the logarithmic gain $\loggain \simeq \de \ln\!N_\mu/\de \lne$ of muons with growing energy. The maximum-likelihood method was validated with a fast realistic simulation of hybrid events and shown to yield unbiased values for $a$ and $b$. The technical aspects will be presented in a separate paper.

%
%
\begin{figure}[t]
\centering
\includegraphics[width=0.47\textwidth,height=7cm]{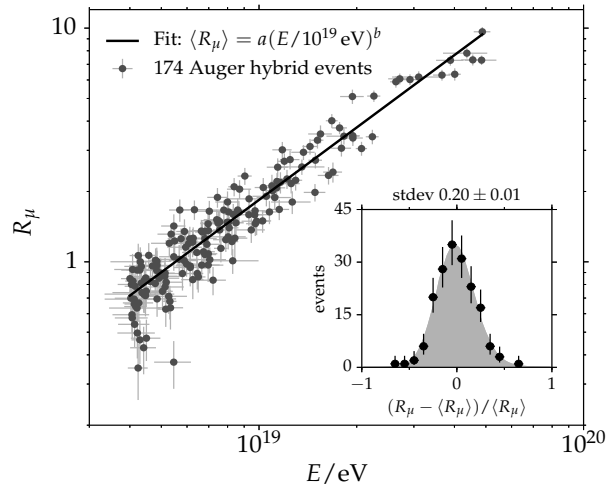}
\caption{The selected hybrid events above \SI{4e18}{eV} and a fit of the power law $\mrmu = a\,\langle E/10^{19}\,\text{eV} \rangle^b$. The error bars indicate statistical detection uncertainties only. The inset shows a histogram of the residuals around the fitted curve (black dots) and for comparison the expected residual distribution computed from the fitted probability model that describes the fluctuations.}
\label{fig:fig03}
\end{figure}
%
%

The data and results of the fit are shown in \fg{fig03}. We obtain
\begin{linenomath}
\begin{align}
a &= \mrmu (10^{19}\,\text{eV}) = (1.841 \pm 0.029 \pm 0.324\,\sys), \\
b &= \loggain = (1.029 \pm 0.024 \pm 0.030\,\sys), \\
& \sigma[R_\mu]/R_\mu = (0.136 \pm 0.015 \pm 0.033\,\sys).
\end{align}
\end{linenomath}
At a zenith angle of $67^\circ$, this corresponds to $(2.68 \pm 0.04 \pm 0.48\,\sys) \times 10^{7}$ muons with energies larger than \SI{0.3}{GeV} that reach \SI{1425}{m} altitude in an average $10^{19}\,\si{eV}$ shower.

The fitted model agrees well with data. To obtain a goodness-of-fit estimator, we compute the histogram of the residuals $(\rmu - \mrmu)/\mrmu$ and compare it with its expectation $g\big((\rmu - \mrmu)/\mrmu \big) = \int g'\big((\rmu - \mrmu(E))/\mrmu(E), E \big) \de E$ computed from the fitted two-dimensional probability density function $f(\rmu, E)$. Histogram and expectation are shown in the inset of \fg{fig03}. The comparison yields a reduced chi-square value $\chi^2/n_\text{dof} = 4.9/10$ for the fitted model.

The systematic uncertainty of the absolute scale $\mrmu (10^{19}\,\si{eV})$ of 18\,\% combines the intrinsic uncertainty of the $\rmu$-measurement (11\,\%) and the uncertainty of the Auger energy scale (14\,\%)~\cite{Verzi2013}. The systematic uncertainty of the logarithmic gain $\loggain$ of 3\,\% is derived from variations of the FD selection cuts (2\,\%), variations of the bias correction of $\rmu$ within its systematic uncertainty (1\,\%), variations of the distribution assumptions on the intrinsic $\rmu$-fluctuations (1\,\%), and by assuming a residual zenith-angle dependence of the ratio $\rmu/E$ that cannot be detected within the current statistics (0.5\,\%). The third parameter $\sigma[R_\mu]/R_\mu$, the relative size of the intrinsic fluctuations, is effectively obtained by subtraction of the detection uncertainties from the total spread. Its systematic uncertainty of $\pm 0.033$ is estimated from the variations just described ($\pm 0.014\,\sys$ in total), and by varying the detection uncertainties within a plausible range ($\pm 0.030\,\sys$).

At $\theta = 67^\circ$, the average zenith angle of the data set, $\rmu = 1$ corresponds to $\nmu = \num{1.455e+07}$ muons at the ground with energies above \SI{0.3}{GeV}. For model comparisons, it is sufficient to simulate showers at this zenith angle down to an altitude of \SI{1425}{m} and count muons at the ground with energies above \SI{0.3}{GeV}. Their number should then be divided by $\nmu =\num{1.455e+07}$ to obtain $\rmumc$, which can be directly compared to our measurement.

Our fit yields the average muon content $\mrmu$. For model comparisons the average logarithmic muon content, $\mlnrmu$, is also of interest, as we will see in the next section. The relationship between the two depends on shape and size of the intrinsic fluctuations. We compute $\mlnrmu$ numerically based on our fitted model of the intrinsic fluctuations:
\begin{linenomath}
\begin{align}
\mlnrmu(10^{19}\,\text{eV}) &= \int_0^\infty \lnrmu \, \mathcal N(\rmu) \, \de \rmu \nonumber \\
&= 0.601 \pm 0.016 \,_{-0.201}^{+0.167}\sys,
\end{align}
\end{linenomath}
where $\mathcal N(\rmu)$ is a Gaussian with mean $\mrmu$ and spread $\sigma[\rmu]$ as obtained from the fit. The deviation of $\mlnrmu$ from $\ln\mrmu$ is only 2\,\% so that the conversion does not lead to a noticeable increase in the systematic uncertainty.

Several consistency checks were performed on the data set. We found no indications for a seasonal variation, nor for a dependence on the zenith angle or the distance of the shower axis to the fluorescence telescopes.

\section{Model comparison and discussion}
%
%
\begin{figure}[t]
\centering
\includegraphics[width=0.47\textwidth,height=7cm]{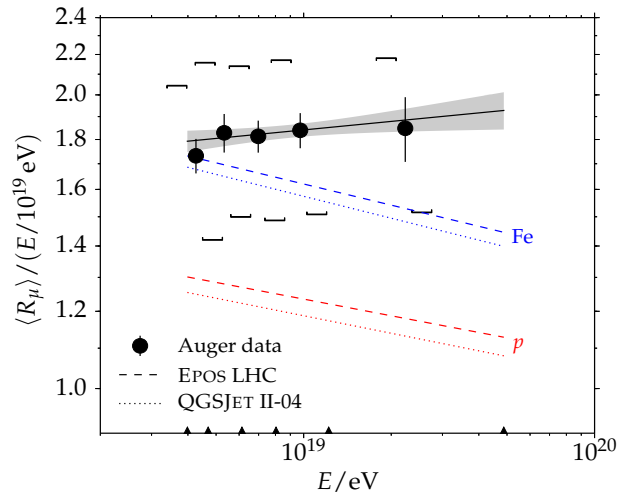}
\caption{Average muon content $\mrmu$ per shower energy $E$ as a function of the shower energy $E$ in double logarithmic scale. Our data is shown bin-by-bin (circles) together with the fit discussed in the previous section (line). Square brackets indicate the systematic uncertainty of the measurement, the diagonal offsets represent the correlated effect of systematic shifts in the energy scale. The grey band indicates the statistical uncertainty of the fitted line. Shown for comparison are theoretical curves for proton and iron showers simulated at $\theta = 67^\circ$ (dotted and dashed lines). Black triangles at the bottom show the energy bin edges. The binning was adjusted by an algorithm to obtain equal numbers of events per bin.}
\label{fig:fig04}
\end{figure}
%
%

A simple comparison of our data with air showers simulated at the mean zenith angle $\theta = 67^\circ$ with the hadronic interaction models \qgsjet II-04 and \epos\ LHC is shown in \fg{fig04}. The ratio $\mrmu/(E/10^{19}\,\text{eV})$ cancels most of the energy scaling, and emphasizes the effect of the cosmic-ray mass $A$ on the muon number. We compute the ratio from \eq{fit} (line), and alternatively by a bin-wise averaging of the original data (data points). The two ways of computing the ratio are visually in good agreement, despite minor bin-to-bin migration effects that bias the bin-by-bin method. The fitting approach we used for the data analysis avoids the migration bias by design.

Proton and iron showers are well separated, which illustrates the power of $\mrmu$ as a composition estimator. A caveat is the large systematic uncertainty on the absolute scale of the measurement, which is mainly inherited from the energy scale~\cite{Verzi2013}. This limits its power as a mass composition estimator, but we will see that our measurement contributes valuable insights into the consistency of hadronic interaction models around and above energies of $10^{19}$\,eV, where other sensitive data are sparse.

A hint of a discrepancy between the models and the data is the high abundance of muons in the data. The measured muon number is higher than in pure iron showers, suggesting contributions of even heavier elements. This interpretation is not in agreement with studies based on the depth of shower maximum~\cite{Letessier-Selvon2013}, which show an average logarithmic mass $\mlna$ between proton and iron in this energy range. We note that our data points can be moved between the proton and iron predictions by shifting them within the systematic uncertainties, but we will demonstrate that this does not completely resolve the discrepancy. The logarithmic gain $\loggain$ of the data is also large compared to proton or iron showers. This suggests a transition from lighter to heavier elements that is also seen in the evolution of the average depth of shower maximum.

We will now quantify the disagreement between model predictions and our data with the help of the mass composition inferred from the average depth $\mxmax$ of the shower maximum. A valid hadronic interaction model has to describe all air shower observables consistently. We have recently published the mean logarithmic mass $\mlna$ derived from the measured average depth of the shower maximum $\mxmax$~\cite{Letessier-Selvon2013}. We can therefore make predictions for the mean logarithmic muon content $\mlnrmu$ based on these $\mlna$ data, and compare them directly to our measurement.

We consider \qgsjet01, \qgsjet II-03, \qgsjet II-04, and \epos\ LHC for this comparison. The relation of $\mxmax$ and $\mlna$ at a given energy $E$ for these models is in good agreement with the prediction from the generalized Heitler model of hadronic air showers
\begin{linenomath}
\begin{equation}\label{eq:xmax_conv}
  \mxmax = \mxmax_p + f_E \mlna,
\end{equation}
\end{linenomath}
where $\mxmax_p$ is the average depth of the shower maximum for proton showers at the given energy and $f_E$ an energy-dependent parameter~\cite{Abreu2013, Ahn2013}. The parameters $\mxmax_p$ and $f_E$ were computed from air shower simulations for each model.

We derive a similar expression from \eq{nmu_heitler} by substituting $N_{\mu,p} = (E/\xi_{\mathrm c})^\beta$ and computing the average logarithm of the muon number
\begin{linenomath}
\begin{align}
  \label{eq:nmu_conv}
  \mlnnmu &= \mlnnmu_p + (1 - \beta) \mlna \\
  \label{eq:beta}
  \beta &= 1 - \frac{\mlnnmu_\text{Fe} - \mlnnmu_p}{\ln 56}.
\end{align}
\end{linenomath}
Since $\nmu \propto \rmu$, we can replace $\lnnmu$ by $\lnrmu$. The same can be done in \eq{slope_heitler}, which also holds for averages due to the linearity of differentiation.

We estimate the systematic uncertainty of the approximate Heitler model by computing $\beta$ from \eq{beta}, and alternatively from $\de \mlnrmu_\text{\it p} / \de \lne$ and $\de \mlnrmu_\text{Fe} / \de \lne$. The three values would be identical if the Heitler model was accurate. Based on the small deviations, we estimate $\sigma_\text{sys}[\beta] = 0.02$. By propagating the systematic uncertainty of $\beta$, we arrive at a small systematic uncertainty for predicted logarithmic muon content of $\sigma_\text{sys}[\mlnrmu] < 0.02$.

%
%
\begin{figure}[t]
\centering
\includegraphics[width=0.47\textwidth,height=7cm]{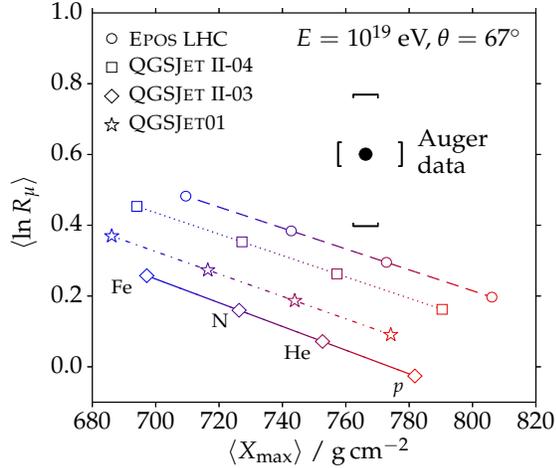}
\caption{Average logarithmic muon content $\mlnrmu$ (this study) as a function of the average shower depth $\mxmax$ (obtained by interpolating binned data from Ref.~\cite{Letessier-Selvon2013}) at $10^{19}$\,eV. Model predictions are obtained from showers simulated at $\theta = 67^\circ$. The predictions for proton and iron showers are directly taken from simulations. Values for intermediate masses are computed with the Heitler model described in the text.}
\label{fig:fig05}
\end{figure}
%
%

With \eq{xmax_conv} and \eq{nmu_conv}, we convert the measured mean depth $\mxmax$ into a prediction of the mean logarithmic muon content $\mlnrmu$ at $\theta = 67^\circ$ for each hadronic interaction model. The relationship between $\mxmax$ and $\mlnrmu$ can be represented by a line, which is illustrated in \fg{fig05}. The Auger measurements at $10^{19}$\,eV are also shown. The discrepancy between data and model predictions is shown by a lack of overlap of the data point with any of the model lines.

%
%
\begin{figure}[t]
\centering
\includegraphics[width=0.47\textwidth,height=7cm]{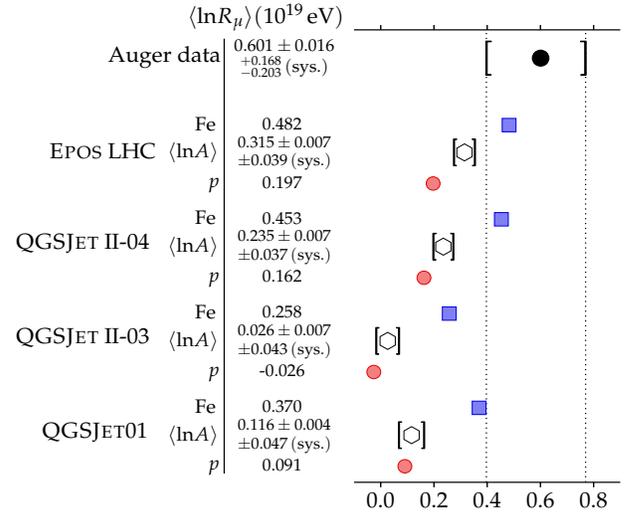}
\caption{Comparison of the mean logarithmic muon content $\mlnrmu$ at $10^{19}$\,eV obtained from Auger data with model predictions for proton and iron showers simulated at $\theta = 67^\circ$, and for such mixed showers with a mean logarithmic mass that matches the mean shower depth $\mxmax$ measured by the FD. Brackets indicate systematic uncertainties. Dotted lines show the interval obtained by adding systematic and statistical uncertainties in quadrature. The statistical uncertainties for proton and iron showers are negligible and suppressed for clarity.}
\label{fig:fig06a}
\end{figure}
%
%

%
%
\begin{figure}[t]
\centering
\includegraphics[width=0.47\textwidth,height=7cm]{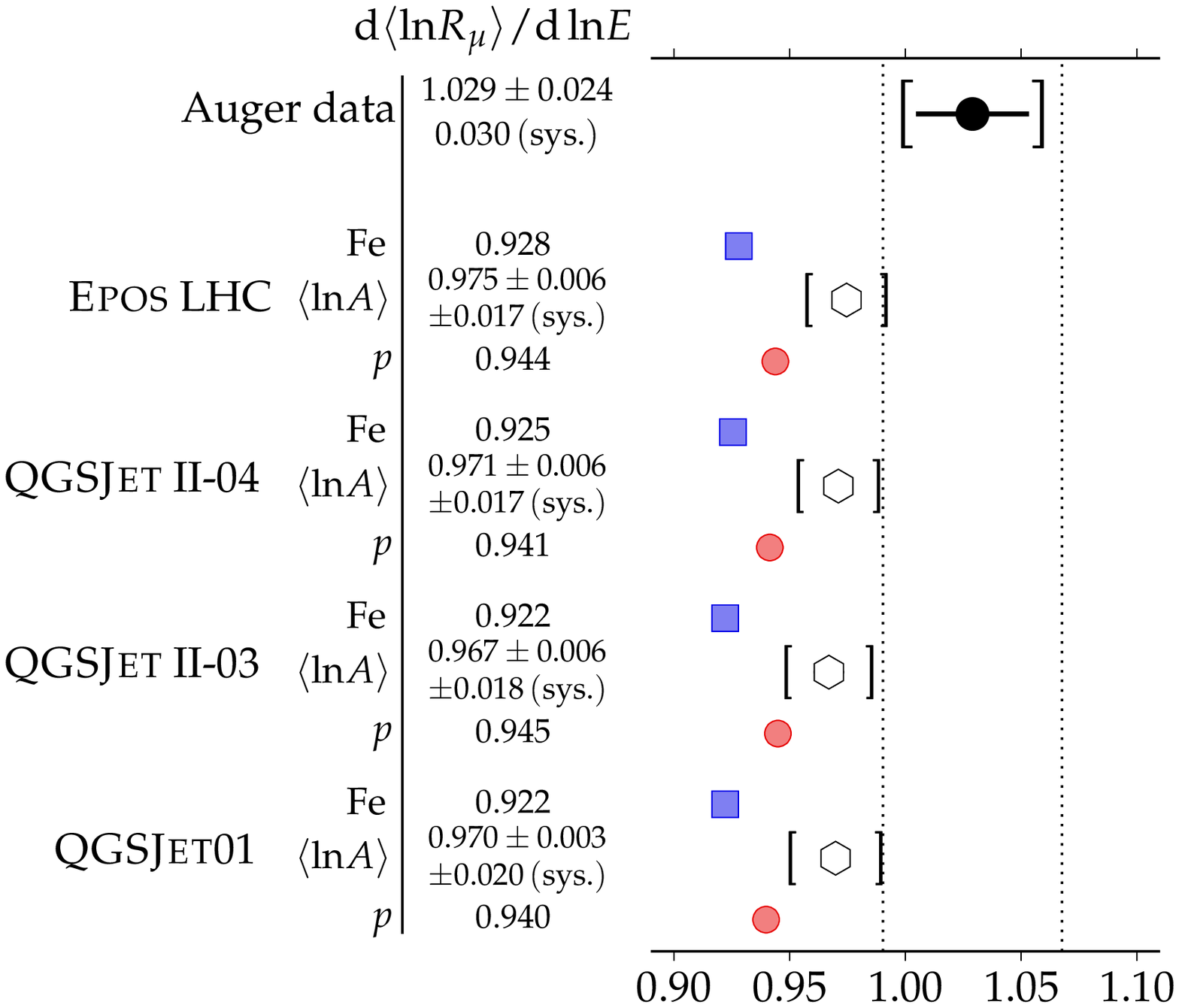}
\caption{Comparison of the logarithmic gain $\loggain$ between \SI{4e18}{eV} and \SI{5e19}{eV} with model predictions in the same style as in \fg{fig06a}.}
\label{fig:fig06b}
\end{figure}
%
%

The model predictions of $\mlnrmu$ and $\loggain$ are summarized and compared to our measurement in \fg{fig06a} and \ref{fig:fig06b}, respectively. For \qgsjet II-03, \qgsjet II-04, and \epos\ LHC, we use estimated $\mlna$ data from Ref.~\cite{Letessier-Selvon2013}. Since \qgsjet01 has not been included in that reference, we compute $\mlna$ using \eq{xmax_conv}~\cite{Abreu2013} from the latest $\mxmax$ data~\cite{Letessier-Selvon2013}. The systematic uncertainty of the $\mlnrmu$ predictions is derived by propagating the systematic uncertainty of $\mlna$ ($\pm 0.03\,\sys$), combined with the systematic uncertainty of the Heitler model ($\pm 0.02\,\sys$). The predicted logarithmic gain $\loggain$ is calculated through \eq{slope_heitler}, while $\de \lna /\de \lne$ is obtained from a straight line fit to $\mlna$ data points between \SI{4e18}{eV} and \SI{5e19}{eV}. The systematic uncertainty of the $\loggain$ predictions is derived by varying the fitted line within the systematic uncertainty of the $\mlna$ data ($\pm 0.02\,\sys$), and by varing $\beta$ within its systematic uncertainty in \eq{slope_heitler} ($\pm 0.005\,\sys$).

The four hadronic interaction models fall short in matching our measurement of the mean logarithmic muon content $\mlnrmu$. \qgsjet II-04 and \epos\ LHC have been updated after the first LHC data. The discrepancy is smaller for these models, and \epos\ LHC performs slightly better than \qgsjet II-04. Yet none of the models is covered by the total uncertainty interval. The minimum deviation is $1.4\,\sigma$. To reach consistency, the mean muon number around $10^{19}\,\si{eV}$ in simulations would have to be increased by 30\,\% to 80\,\% $^{+17}_{-20}\,\sys\,\%$. If on the other hand the predictions of the latest models were close to the truth, the Auger energy scale would have to be increased by a similar factor to reach agreement. Without a self-consistent description of air shower observables, conclusions about the mass composition from the measured absolute muon content remain tentative.

The situation is better for the logarithmic gain $\loggain$. The measured value is higher than the predictions from $\mlna$ data, but the discrepancy is smaller. If all statistical and systematic uncertainties are added in quadrature, the deviation between measurement and $\mlna$-based predictions is 1.3 to 1.4 $\sigma$. The statistical uncertainty is not negligible, which opens the possibility that the apparent deviation is a statistical fluctuation. If we assume that the hadronic interaction models reproduce the logarithmic gain of real showers, which is supported by the internal consistency of the predictions, the large measured value of $\loggain$ disfavors a pure composition hypothesis. If statistical and systematic uncertainties are added in quadrature, we observe deviations from a pure proton (iron) composition of $2.2\,\sigma$ ($2.6\,\sigma$).

\section{Conclusions}
We presented the first measurement of the mean muon number in inclined air showers with $\theta > 62^\circ$ between \SI{4e18}{eV} and \SI{5e19}{eV} and its logarithmic gain with energy, based on data from a hybrid detector. We explored the sensitivity of the muon number to the cosmic-ray mass composition and challenged predictions of the muon number from hadronic interaction models. 
We observe a muon deficit in simulations of 30\,\% to 80\,\% $^{+17}_{-20}\,\sys\,\%$ at $10^{19}\,\si{eV}$, depending on the model. The estimated deficit takes the mass composition of cosmic rays into account, by comparing our measurement to the average muon number in simulated air showers which match the average depth of shower maximum observed in the data.

Model predictions of the logarithmic gain of muons with rising energy are within the uncertainties compatible with the measured value. The high gain of muons favors a transition from lighter to heavier elements in the considered energy range. The hypothesis of a constant proton composition, supported by measurements of the depth of shower maximum by the Telescope Array~\cite{Allen2013} in the northern hemisphere, is disfavored with respect to our result at the level of $2.2\,\sigma$.


Our measurement of the muon number combined with measurements of the depth of shower maximum provides important insights into the consistency of hadronic interaction models. The hadronic and muonic components of air showers are less well understood than the electromagnetic component, but all three are physically connected. Improvements in the description of the muonic component will also reduce the systematic uncertainty in the simulation of the other components.

This result is compatible with those of independent studies for showers with $\theta < 60^\circ$~\cite{Kegl2013}, in which different methods have been used to derive the fraction of the signal due to muons at 1000~m from the shower core using the temporal distribution of the signals measured with the SD array.

We have demonstrated how the mass composition of cosmic rays can be inferred from the muon number measured at the ground. To fully explore this potential, the apparent muon deficit in air-shower simulations needs to be resolved and the uncertainty of the muon measurement further reduced. The main contributions are the systematic uncertainties in the simulated response of the Auger SD to inclined muons, and the systematic uncertainty in the absolute energy scale. We expect to reduce both of them in the future, which will significantly enhance the constraining power of the muon measurement on the mass composition.


\section{Acknowledgements}
The successful installation, commissioning, and operation of the Pierre Auger Observatory would not have been possible without the strong commitment and effort from the technical and administrative staff in Malarg\"{u}e. 

We are very grateful to the following agencies and organizations for financial support: 
Comisi\'{o}n Nacional de Energ\'{\i}a At\'{o}mica, Fundaci\'{o}n Antorchas, Gobierno De La Provincia de Mendoza, Municipalidad de Malarg\"{u}e, NDM Holdings and Valle Las Le\~{n}as, in gratitude for their continuing cooperation over land access, Argentina; the Australian Research Council; Conselho Nacional de Desenvolvimento Cient\'{\i}fico e Tecnol\'{o}gico (CNPq), Financiadora de Estudos e Projetos (FINEP), Funda\c{c}\~{a}o de Amparo \`{a} Pesquisa do Estado de Rio de Janeiro (FAPERJ), S\~{a}o Paulo Research Foundation (FAPESP) Grants \# 2010/07359-6, \# 1999/05404-3, Minist\'{e}rio de Ci\^{e}ncia e Tecnologia (MCT), Brazil; MSMT-CR LG13007, 7AMB14AR005, CZ.1.05/2.1.00/03.0058 and the Czech Science Foundation grant 14-17501S, Czech Republic;  Centre de Calcul IN2P3/CNRS, Centre National de la Recherche Scientifique (CNRS), Conseil R\'{e}gional Ile-de-France, D\'{e}partement Physique Nucl\'{e}aire et Corpusculaire (PNC-IN2P3/CNRS), D\'{e}partement Sciences de l'Univers (SDU-INSU/CNRS), Institut Lagrange de Paris, ILP LABEX ANR-10-LABX-63, within the Investissements d'Avenir Programme  ANR-11-IDEX-0004-02, France; Bundesministerium f\"{u}r Bildung und Forschung (BMBF), Deutsche Forschungsgemeinschaft (DFG), Finanzministerium Baden-W\"{u}rttemberg, Helmholtz-Gemeinschaft Deutscher Forschungszentren (HGF), Ministerium f\"{u}r Wissenschaft und Forschung, Nordrhein Westfalen, Ministerium f\"{u}r Wissenschaft, Forschung und Kunst, Baden-W\"{u}rttemberg, Germany; Istituto Nazionale di Fisica Nucleare (INFN), Ministero dell'Istruzione, dell'Universit\`{a} e della Ricerca (MIUR), Gran Sasso Center for Astroparticle Physics (CFA), CETEMPS Center of Excellence, Italy; Consejo Nacional de Ciencia y Tecnolog\'{\i}a (CONACYT), Mexico; Ministerie van Onderwijs, Cultuur en Wetenschap, Nederlandse Organisatie voor Wetenschappelijk Onderzoek (NWO), Stichting voor Fundamenteel Onderzoek der Materie (FOM), Netherlands; National Centre for Research and Development, Grant Nos.ERA-NET-ASPERA/01/11 and ERA-NET-ASPERA/02/11, National Science Centre, Grant Nos. 2013/08/M/ST9/00322, 2013/08/M/ST9/00728 and HARMONIA 5 - 2013/10/M/ST9/00062, Poland; Portuguese national funds and FEDER funds within COMPETE - Programa Operacional Factores de Competitividade through Funda\c{c}\~{a}o para a Ci\^{e}ncia e a Tecnologia, Portugal; Romanian Authority for Scientific Research ANCS, CNDI-UEFISCDI partnership projects nr.20/2012 and nr.194/2012, project nr.1/ASPERA2/2012 ERA-NET, PN-II-RU-PD-2011-3-0145-17, and PN-II-RU-PD-2011-3-0062, the Minister of National  Education, Programme for research - Space Technology and Advanced Research - STAR, project number 83/2013, Romania; Slovenian Research Agency, Slovenia; Comunidad de Madrid, FEDER funds, Ministerio de Educaci\'{o}n y Ciencia, Xunta de Galicia, European Community 7th Framework Program, Grant No. FP7-PEOPLE-2012-IEF-328826, Spain; Science and Technology Facilities Council, United Kingdom; Department of Energy, Contract No. DE-AC02-07CH11359, DE-FR02-04ER41300, DE-FG02-99ER41107 and DE-SC0011689, National Science Foundation, Grant No. 0450696, The Grainger Foundation, USA; NAFOSTED, Vietnam; Marie Curie-IRSES/EPLANET, European Particle Physics Latin American Network, European Union 7th Framework Program, Grant No. PIRSES-2009-GA-246806; and UNESCO.

\bibliography{auger,local}
\bibliographystyle{prl}

\end{document}